%% file: main.tex
\documentclass[conference]{IEEEtran}
\IEEEoverridecommandlockouts

\usepackage{url}
\usepackage[colorlinks,allcolors=blue]{hyperref}

\usepackage{cite}

\usepackage{amsmath,amsfonts}
\usepackage{mathtools}
\usepackage{graphicx}
\usepackage{textcomp}
\usepackage{xcolor}
\usepackage{pgfplots}
\usepackage[makeroom]{cancel}

\usepackage{algorithm}
\usepackage{algpseudocode}

\pgfplotsset{compat=newest}
\usetikzlibrary{spy,backgrounds,bayesnet}
\usepgfplotslibrary{units}
\usepackage{pgfplotstable}

\ifCLASSOPTIONcompsoc
  \usepackage[caption=false,font=normalsize,labelfont=sf,textfont=sf]{subfig}
\else
  \usepackage[caption=false,font=footnotesize]{subfig}
\fi

\DeclareMathOperator*{\argmin}{arg\,min}
\newcommand{\vx}{\mathbf{x}}
\newcommand{\vy}{\mathbf{y}}
\newcommand{\vm}{\mathbf{m}}
\newcommand{\vP}{\mathbf{P}}
\newcommand{\vQ}{\mathbf{Q}}
\newcommand{\vR}{\mathbf{R}}
\newcommand{\vf}{\mathbf{f}}
\newcommand{\vh}{\mathbf{h}}
\newcommand{\vq}{\mathbf{q}}
\newcommand{\vr}{\mathbf{r}}
\newcommand{\vA}{\mathbf{A}}
\newcommand{\vF}{\mathbf{F}}
\newcommand{\vH}{\mathbf{H}}

\newcommand{\ve}{\mathbf{e}}
\newcommand{\vc}{\mathbf{c}}
\newcommand{\vb}{\mathbf{b}}
\newcommand{\vK}{\mathbf{K}}
\newcommand{\vG}{\mathbf{G}}

\newcommand{\vU}{\mathbf{U}}

\newcommand{\hvx}{\hat{\mathbf{x}}}

\newcommand{\tL}{\tilde{L}}

\newcommand{\vPsi}{\mathbf{\Psi}}
\newcommand{\vGamma}{\mathbf{\Gamma}}
\newcommand{\vPhi}{\mathbf{\Phi}}
\newcommand{\vI}{\mathbf{I}}
\newcommand{\vp}{\mathbf{p}}

\newcommand{\vOmega}{\mathbf{\Omega}}
\newcommand{\vtau}{\mathbf{\tau}}

\def\BibTeX{{\rm B\kern-.05em{\sc i\kern-.025em b}\kern-.08em
    T\kern-.1667em\lower.7ex\hbox{E}\kern-.125emX}}

\begin{document}

\title{A Recursive Newton Method for Smoothing \\ in Nonlinear State Space Models%
\thanks{This work was funded by the Academy of Finland\textsuperscript{*} and the Finnish Center for Artificial Intelligence (FCAI)\textsuperscript{\textdagger}.}
}

\author{
Fatemeh Yaghoobi\textsuperscript{*}, Hany Abdulsamad\textsuperscript{\textdagger}, Simo Särkkä \\
Department of Electrical Engineering and Automation, Aalto University, Finland \\
\{fatemeh.yaghoobi, hany.abdulsamad, simo.sarkka\}@aalto.fi
}

\maketitle

\begin{abstract}
In this paper, we use the optimization formulation of nonlinear Kalman filtering and smoothing problems to develop second-order variants of iterated Kalman smoother (IKS) methods. We show that Newton's method corresponds to a recursion over affine smoothing problems on a modified state-space model augmented by a pseudo measurement. The first and second derivatives required in this approach can be efficiently computed with widely available automatic differentiation tools. Furthermore, we show how to incorporate line-search and trust-region strategies into the proposed second-order IKS algorithm in order to regularize updates between iterations. Finally, we provide numerical examples to demonstrate the method's efficiency in terms of runtime compared to its batch counterpart. 
\end{abstract}


\section{Introduction}
State estimation problem in nonlinear state-space models (SSMs) plays an important role in various areas of applications such as in control theory, signal processing, and robotics \cite{sarkka2013bayesian, Bar-Shalom+Li+Kirubarajan:2001, bar1995multitarget}. In this paper, we are interested in solving state estimation problems in SSMs of the form
\begin{equation} \label{eq:SSM-nonlinear/Gaussian}
\begin{split}
  \vx_{k} &= \vf(\vx_{k-1}) + \vq_{k-1}, \quad 
  \vy_{k} = \vh(\vx_{k}) + \vr_{k},
\end{split}
\end{equation}
$\vx_{k} \in \mathbb{R}^d$ is the state at time step $k$, $\vy_{k} \in \mathbb{R}^m$ is the measurement at the same time step, $\vf(.)$ is the state transition function, and $\vh(.)$ is the observation function. Furthermore, $\vq_{k}$ and $\vr_{k}$ are the process and measurement noises, assumed to be Gaussian with zero mean and covariance matrices $\vQ$ and $\vR$, respectively. The prior distribution of the state at $k=0$ is Gaussian with known mean $\vm_{0}$ and covariance $\vP_{0}$.

The smoothing problem (see, e.g., \cite{sarkka2013bayesian}) amounts to computing the estimate of the state $\vx_k$ given a batch of measurements $\vy_{1}, \ldots, \vy_{N}$, where $k \in \{0,\ldots,N\}$. The Kalman filter \cite{kalman1960new} and Rauch--Tung--Striebel (RTS) smoother \cite{rauch1965maximum} for linear SSM and their extension for nonlinear systems (see, e.g., \cite{Bar-Shalom+Li+Kirubarajan:2001, sarkka2013bayesian, jazwinski2007stochastic, julier2000new, bell1994iterated, Garcia+Svensson+Moreland+Sarkka:2015, Garcia:2017, Tronarp+Fernandez+Sarkka:2018}) provide powerful recursive solutions which are optimal in the minimum mean squared error (MMSE) sense. 

On the other hand, the smoothing problem can be viewed in an optimization framework (see, e.g., \cite{bell1994iterated, sarkka2020levenberg}), where the aim is to find the maximum a posteriori (MAP) trajectory estimate, that is, the trajectory $\vx^*_{0:N}$ which maximizes $p(\vx_{0:N} \mid \vy_{1:N})$. 

For the SSM of the form~\eqref{eq:SSM-nonlinear/Gaussian}, the MAP estimate is the minimizer of the negative log-posterior 
\begin{equation}
    \vx^*_{0:N} = \argmin_{\vx_{0:N}}{\, \, L(\vx_{0:N})},
\end{equation}
where the negative log-posterior is given by
\begin{align}
    & L(\vx_{0:N}) = \frac{1}{2} \lVert \vx_{0}
    - \vm_{0} \rVert^{2}_{\vP_{0}^{-1}} + \frac{1}{2} \sum_{k=1}^{N} \lVert \vx_{k} - \vf(\vx_{k-1}) \rVert^{2}_{\vQ^{-1}} \notag \\
    & \quad + \frac{1}{2} \sum_{k=1}^{N} \lVert \vy_{k} - \vh(\vx_{k}) \rVert^{2}_{\vR^{-1}}, ~\, \, \small{\text{with}} \, \, \tiny{\lVert \vx \rVert^{2}_{\vA} \coloneqq \vx^\top \vA \vx}. \label{eq:objective-function}
\end{align}
Viewing the state estimation problem from an optimization standpoint enables us to employ several optimization techniques \cite{wright1999numerical}. One widely-used example in the filtering and smoothing literature is the Gauss--Newton (GN) method \cite{garcia2014gaussian, fatemi2012study, moriyama2003incremental}, which has a close relationship with iterated extended Kalman filtering and smoothing methods \cite{Bell:1993, bell1994iterated}. In particular, for nonlinear SSM with additive Gaussian noise, Bell \cite{bell1994iterated} proved that the GN-method is equivalent to the iterated extended Kalman smoother (IEKS), a recursive method with less computational complexity than batch GN-methods. Recently, S\"arkk\"a \& Svensson \cite{sarkka2020levenberg} developed line-search and Levenberg-Marquart extensions of the IEKS method.

Newton's method has received less attention as an optimization method to solve smoothing problems due to the effort associated with computing second-order derivatives. However, the availability of automatic differentiation tools has eliminated the need for manual computation, making Newton's method attractive for smoothing problems. Although the application of Newton's method to filtering and smoothing has been mentioned in literature \cite{ humpherys2012fresh, skoglund2015extended, ollivier2019extended}, the full Newton version of the IKS is yet to be realized.

The contribution of this paper is to develop the Newton formulation of iterated Kalman smoothers while leveraging automatic differentiation tools to compute the derivatives and Hessians. We also present robust modifications of the proposed method that incorporate line-search and trust-region schemes into the recursive structure.

This paper is structured as follows: Section~\ref{sec:Newton-IKS} presents Newton's method for the MAP problem in batch and recursive form. Section~\ref{sec:ls_tr_newton} presents line-search and trust-region strategies to enhance the robustness of iterative Newton updates. Section \ref{sec:experiment} analyzes the efficiency of the proposed recursive methods in the sense of runtime on a numerical example.

\section{Newton Iterated Kalman Smoother} \label{sec:Newton-IKS}
Assuming a SSM of the form~\eqref{eq:SSM-nonlinear/Gaussian} and the objective specified in Equation~\eqref{eq:objective-function}, our aim, in this section, is to use Newton's optimization technique to minimize the objective function and develop the corresponding batch solution. Subsequently, we present a recursive alternative analogous to the IKS to improve computational efficiency.

\subsection{Batch Newton Optimization} \label{sec:batch_newton}
The batch solution for smoothing follows the standard iterative optimization framework without specifically leveraging the underlying temporal structure of the problem. Accordingly, we can implement Newton's method as a generic second-order optimization of Equation~\eqref{eq:objective-function} with respect to a decision variable $\vx_{0:N}$ with a dimension $d_N = d\times N$.

At every iteration $i$, Newton's method approximates a twice differentiable objective $L(\vx_{0:N})$ up to the second order in the neighborhood of a nominal trajectory $\hvx^{(i)}_{0:N}$
\begin{align} \label{eq:quadratic-approximation} 
    L(\vx_{0:N}) \approx \,
    & L(\hvx^{(i)}_{0:N}) + \nabla L^\top(\hvx^{(i)}_{0:N}) (\vx_{0:N} - \hvx^{(i)}_{0:N}) \\ 
    & + \frac{1}{2}(\vx_{0:N} - \hvx^{(i)}_{0:N})^\top \nabla^{2}L(\hvx^{(i)}_{0:N} )(\vx_{0:N} - \hvx^{(i)}_{0:N}), \notag
\end{align}
where $\nabla L(.)$ and $\nabla^{2}L(.)$ denote the gradient and the Hessian of $L(.)$, respectively. Using this quadratic approximation, we get the Newton update rule
\begin{equation} \label{eq:Newton-update}
    \hvx^{(i+1)}_{0:N} = \hvx^{(i)}_{0:N} - (\nabla^{2}L(\hvx^{(i)}_{0:N}) + \lambda \, \vI_{d_N})^{-1} \, \nabla L(\hvx^{(i)}_{0:N}).
\end{equation}
Note that we have included a diagonal regularization term $\lambda \, \vI_{d_N}$, with $\lambda \geq 0$, to ensure a positive-definite Hessian and a valid descent direction.

Despite the convenience of automatic differentiation frameworks that readily deliver $\nabla L(.)$ and $\nabla^{2}L(.)$, the computational effort associated with the Newton update in Equation~\eqref{eq:Newton-update} is still a major issue. The Hessian $\nabla^{2}L(.)$ is of dimensions $d_N  \times d_N$, and its inversion leads to a worst-case computational complexity $\mathcal{O}(N^{3}d^{3})$, which scales poorly both in the state dimension and the trajectory length.

In the following, we will rely on the quadratic approximation in Equation~\eqref{eq:quadratic-approximation}. However, by taking advantage of the temporal structure of the state-space model, we will construct a modified affine state-space model and derive a recursive algorithm akin to the iterated Kalman smoother, leading to a considerable reduction in computational complexity.

\subsection{Recursive Newton Optimization}\label{sec:recursive-Newton-IKS}
Constructing the modified state-space model requires analyzing the first- and second-order approximations of the individual terms in Equation~\eqref{eq:objective-function}. We start by considering the approximation of the transition dynamics term. For convenience, we define
\begin{equation*}
    S(\vx_{0:N}) \coloneqq \sum_{k=1}^{N} S_{k}(\vx_{k}, \vx_{k-1}) = \sum_{k=1}^{N} \lVert \vx_{k} - \vf(\vx_{k-1}) \rVert^{2}_{\vQ^{-1}},
\end{equation*}
and expand it around the current nominal trajectory $\hvx^{(i)}_{0:N}$. For a simplified notation, we drop the iteration index $i$
\begin{equation} \label{eq:tranition_expansion}
    S(\vx_{0:N}) \approx
    \begin{aligned}[t]
        & \frac{1}{2} \delta {\vx}_{0:N}^{\top} \, \nabla^{2} S(\hvx_{0:N}) \, \delta {\vx}_{0:N} \\
        & + \nabla S^\top(\hvx_{0:N}) \, \delta {\vx}_{0:N} + S(\hvx_{0:N}),
    \end{aligned}
\end{equation}
where $\delta {\vx}_{0:N} = \vx_{0:N} - \hvx_{0:N}$ and
\begin{align} 
    & \nabla S^{\top}(\hvx_{0:N}) \, \delta {\vx}_{0:N} =
    2 \sum_{k=1}^{N} (\hvx_{k} - \vf(\hvx_{k-1}))^{\top} \vQ^{-1} \, \delta {\vx}_{k} \label{eq:first_order_transition} \\
    & \quad \quad \quad - 2 \sum_{k=1}^{N} (\hvx_{k} - \vf(\hvx_{k-1}))^{\top} \vQ^{-1} \vF_{\vx}(\hvx_{k-1}) \, \delta {\vx}_{k-1}, \notag  \\
    & \frac{1}{2} \delta {\vx}_{0:N}^{\top} \, \nabla^{2} S(\hvx_{0:N}) \, \delta {\vx}_{0:N} = \sum_{k=1}^{N} \delta \vx_{k}^{\top} \vQ^{-1} \delta \vx_{k} \label{eq:second_order_transition} \\
    & \quad \quad \quad - 2 \sum_{k=1}^{N} \delta \vx_{k}^{\top} \vF^{\top}_{\vx}(\hvx_{k-1}) \, \vQ^{-1} \delta \vx_{k-1} \notag \\ 
    & \quad \quad \quad + \sum_{k=1}^{N} \delta \vx_{k-1}^{\top} \vF^{\top}_{\vx}(\hvx_{k-1}) \, \vQ^{-1} \vF_{\vx}(\hvx_{k-1}) \, \delta \vx_{k-1} \notag \\
    & \quad \quad \quad - \sum_{k=1}^{N} \delta \vx_{k-1}^{\top} \vF^{\top}_{\vx \vx}(\hvx_{k-1}) \cdot \vQ^{-1} (\hvx_{k} - \vf(\hvx_{k-1})) \, \delta \vx_{k-1}, \notag
\end{align}
where $\vF_{\vx}(.)$ is the Jacobian and $\vF_{\vx \vx}(.)$ is a third-rank Hessian tensor of the transition function $\vf(.)$. The notation $(M \cdot v)$ refers to a tensor dot product so that $(M \cdot v)_{ij} = \sum_{k} M_{ijk} v_{k}$.

Plugging Equations~\eqref{eq:first_order_transition} and \eqref{eq:second_order_transition} into Equation~\eqref{eq:tranition_expansion} and applying simple algebraic manipulations, we arrive at the following decomposition of the quadratic expansion in Equation~\eqref{eq:tranition_expansion}
\begin{equation} 
    \begin{split}
    S(\vx_{0:N}) \approx
        & \sum_{k=1}^{N} \lVert \vx_{k} - \vF_{k-1} \, \vx_{k-1} - \vb_{k-1} \rVert^{2}_{\vQ^{-1}} \\
        & + \sum_{k=1}^{N} \lVert \hvx_{k-1} - \vx_{k-1} \rVert^{2}_{\vPsi_{k-1}},
    \end{split}
    \label{eq:modified_transition}
\end{equation}
where
\begin{gather*}
    \vF_{k-1} = \vF_{\vx}(\hvx_{k-1}), \\
    \vb_{k-1} = \vf(\hvx_{k-1}) -  \vF_{\vx}(\hvx_{k-1}) \, \hvx_{k-1}, \\
    \vPsi_{k-1} = - \vF^{\top}_{\vx \vx}(\hvx_{k-1}) \cdot \vQ^{-1} (\hvx_{k} - \vf(\hvx_{k-1})).
\end{gather*}

A similar second-order expansion can be carried out for the observation model term in Equation~\eqref{eq:objective-function}. Again, for convenience, we define the following
\begin{equation*}
    G(\vx_{0:N}) \coloneqq \sum_{k=1}^{N} G_{k}(\vx_{k}) = \sum_{k=1}^{N} \lVert \vy_{k} - \vh(\vx_{k}) \rVert^{2}_{\vR^{-1}},
\end{equation*}
and expand it to the second order around $\hvx_{0:N}$
\begin{equation} \label{eq:observation_expansion}
    G(\vx_{0:N}) \approx
    \begin{aligned}[t]
        & \frac{1}{2} \delta {\vx}_{0:N}^{\top} \, \nabla^{2} G(\hvx_{0:N}) \, \delta {\vx}_{0:N} \\
        & + \nabla G^{\top}(\hvx_{0:N}) \, \delta {\vx}_{0:N} + G(\hvx_{0:N}), &
    \end{aligned}
\end{equation}
where the linear and quadratic terms are
\begin{align*} 
    & \nabla G^{\top}(\hvx_{0:N}) \, \delta {\vx}_{0:N} = - 2 \sum_{k=1}^{N} (\vy_{k} - \vh(\hvx_{k}))^{\top} \vR^{-1} \vH_{\vx}(\hvx_{k}) \, \delta \vx_{k}, \\
    & \frac{1}{2} \delta {\vx}_{0:N}^{\top} \, \nabla^{2} G(\hvx_{0:N}) \, \delta {\vx}_{0:N} \! = \! \sum_{k=1}^{N} \delta \vx_{k}^{\top} \vH_{\vx}^{\top}(\hvx_{k}) \, \vR^{-1} \vH_{\vx}(\hvx_{k}) \delta \vx_{k} \\
    & \qquad \qquad \qquad \quad - \sum_{k=1}^{N} \delta \vx_{k}^{\top} \vH_{\vx \vx}^{\top}(\hvx_{k}) \cdot \vR^{-1} (\vy_{k} - \vh(\hvx_{k})) \, \delta \vx_{k}.
\end{align*}
The matrix $\vH_{\vx}(.)$ is the Jacobian and $\vH_{\vx \vx}(.)$ is a third-rank Hessian tensor of the observation function $\vh(.)$. Similarly, by rearranging these terms, we can construct a specific decomposition of the quadratic expansion in Equation~\eqref{eq:observation_expansion}
\begin{equation} \label{eq:modified_observation}
    G(\vx_{0:N}) \! \approx \sum_{k=1}^{N} \lVert \vy_{k} - \vH_{k} \, \vx_{k} - \vc_{k} \rVert^{2}_{\vR^{-1}} + \! \sum_{k=1}^{N} \lVert \hvx_{k} - \vx_{k} \rVert^{2}_{\vGamma_{k}},
\end{equation}
where
\begin{gather*}
    \vH_{k} = \vH_{\vx}(\hvx_{k}), \\
    \vc_{k} = \vh(\hvx_{k}) - \vH_{\vx}(\hvx_{k}) \, \hvx_{k}, \\
    \vGamma_{k} = - \vH_{\vx \vx}^{\top}(\hvx_{k}) \cdot \vR^{-1} (\vy_{k} - \vh(\hvx_{k})).
\end{gather*}
We can now take the second-order terms of the transition and observation functions in Equations~\eqref{eq:modified_transition} and \eqref{eq:modified_observation} and plug them back into the objective in Equation~\eqref{eq:objective-function} which leads to the overall (regularized) second-order approximation
\begin{align} 
    \label{eq:approx-objective-function}
    & \tL(\vx_{0:N}) = \,
    \frac{1}{2} \lVert \vx_{0} - \vm_{0} \rVert^{2}_{{\vP}_{0}^{-1}} + \frac{1}{2} \lVert \vx_{0} - \hvx_{0} \rVert^{2}_{{\vPhi}_{0}^{-1}} \notag \\
    & \qquad + \frac{1}{2} \sum_{k=1}^{N} \lVert \hvx_{k} - \vx_{k} \rVert^{2}_{\vPhi_{k}^{-1}}
    + \frac{1}{2} \sum_{k=1}^{N} \lVert \vy_{k} - \vH_{k} \, \vx_{k} - \vc_{k} \rVert^{2}_{\vR^{-1}} \notag \\
    & \qquad + \frac{1}{2} \sum_{k=1}^{N} \lVert \vx_{k}- \vF_{k-1} \, \vx_{k-1} - \vb_{k-1} \rVert^{2}_{\vQ^{-1}},
\end{align}
where 
\begin{gather*}
    \vPhi_{0} = (\mathbf{\Psi}_{0} + \lambda \, \vI_{d})^{-1}, \\
    \vPhi_{k} = (\vPsi_{k} + \vGamma_{k} + \lambda \, \vI_d)^{-1}, \\ 
    \vPhi_{N} = (\vGamma_{N} + \lambda \, \vI_{d})^{-1}.
\end{gather*}

The result in Equation~\eqref{eq:approx-objective-function} indicates that the second-order approximation of $L(.)$ can be viewed as a first-order approximation of the functions $\vf$ and $\vh$, augmented by an affine pseudo observation model, in which the expansion point $\hvx_{k}$ acts as a pseudo measurement of the state $\vx_{k}$. This interpretation of \eqref{eq:approx-objective-function} corresponds to the \emph{modified} state-space model of the form
\begin{align*}
    \vx_{k} & \approx \vF_{k-1} \, \vx_{k-1} + \vb_{k-1} + \vq_{k},&&  \, \vq_{k} \sim \mathcal{N}(0, \vQ), \\
    \vy_{k} & \approx \vH_{k} \, \vx_{k} + \vc_{k} + \vr_{k},&&  \, \vr_{k} \sim \mathcal{N}(0, \vR), \\
    \hvx_{k} & \approx \vx_{k} + \ve_{k},&& \, \ve_{k} \sim \mathcal{N}(0, \vPhi_{k}),
\end{align*}
with a \emph{modified} prior distribution $\vx_{0} \sim \mathcal{N}(\vtau_{0}, \vOmega_{0})$
\begin{gather*}
    \vOmega_{0} = (\vP^{-1}_{0} + \vPhi^{-1}_{0})^{-1}, \\
    \vtau_{0} = (\vP^{-1}_{0} + \vPhi^{-1}_{0})^{-1} \, (\vP^{-1}_{0} \vm_0 + \vPhi^{-1}_{0} \hvx_0).
\end{gather*}
Note that we have again included a diagonal term $\lambda \, \vI_{d}$ equivalent to that in Section~\ref{sec:batch_newton}. In this modified state-space model, $\lambda \, \vI_{d}$ can be interpreted as regularization of the pseudo observation model to guarantee a positive-definite covariance and well-defined Gaussian noise. The significance of this regularization will become clear in the upcoming section.

Given this modified affine state-space model, we can iteratively minimize the approximate objective in Equation~\eqref{eq:approx-objective-function} by implementing a recursive RTS smoother \cite{rauch1965maximum} that incorporates the pseudo measurements and dramatically lowers the computational complexity to $\mathcal{O}(Nd^{3})$. Algorithm~\ref{alg:Newton-IKS-additive} summarizes a single iteration of a Newton iterated Kalman smoother (Newton-IKS). For more details on smoothing algorithms for affine state space models, we refer to \cite{sarkka2013bayesian}.

\begin{algorithm}[t]
    \caption{One Iteration of the (Regularized) Newton-IKS}\label{alg:Newton-IKS-additive}
    \begin{algorithmic}[1]
        \item \textbf{input:} Nominal trajectory $\hvx^{(i)}_{0:N}$, measurements $\vy_{1:N}$, \,
        Jacobians at nominal: $\vF_{0:N-1}, \, \vH_{1:N}$, offsets at nominal: $\vb_{0:N-1}, \vc_{1:N}$, covariances at nominal: $\vQ, \vR, \vPhi_{1:N}$, prior at nominal: $\vtau_0, \, \vOmega_0$,  and optional regularization $\lambda$
        \vspace{2pt}
        \item \textbf{output:} Smoothed trajectory $\hvx_{0:N}$
        \vspace{2pt}
        \Procedure{Newton-IKS}{$\hvx^{(i)}_{0:N}, \lambda$}:
        \vspace{2pt}
        \State Set $\vx^{f}_{0} \gets \vtau_{0}(\lambda)$, $\vP^{f}_{0} \gets \vOmega_0(\lambda)$
        \Comment{Initialize}
        \For{$k \gets 1$ \textbf{to} $N$}
            \State $\vx^{p}_{k} \gets \vF_{k-1} \, \vx^{f}_{k-1} + \vb_{k-1}$ \Comment{Prediction}
            \vspace{2pt}
            \State $\vP^{p}_{k} \gets \vF_{k-1} \vP^{f}_{k-1} \vF^\top_{k-1} + \vQ$
            \vspace{5pt}
            \State $\mathbf{\mu}_{k} \gets \vH_{k} \, \vx^{p}_{k} + \vc_{k}$
            \vspace{2pt}
            \State $\mathbf{\Sigma}_{k} \gets \vH_{k} \, \vP^{p}_{k} \, \vH^\top_{k} + \vR$
            \vspace{2pt}
            \State $\vK_{k} \gets \vP^{p}_{k} \, \vH^\top_{k} \mathbf{\Sigma}_{k}^{-1}$
            \vspace{2pt}
            \State $\vx^{y}_{k} \gets \vx^{p}_{k} + \vK_{k} (\vy_{k} - \mathbf{\mu}_{k})$ \Comment{Measure. Update}
            \vspace{2pt}
            \State $\vP^{y}_{k} \gets \vP^{p}_{k} - \vK_{k} \mathbf{\Sigma}_{k} \vK_{k}^\top$
            \vspace{5pt}
            \State $\mathbf{\Delta}_{k} \gets \vP^{y}_{k} + \vPhi_{k}(\lambda)$
            \vspace{2pt}
            \State $\vU_{k} \gets \vP^{y}_{k} \mathbf{\Delta}_{k}^{-1}$
            \vspace{2pt}
            \State $\vx^{f}_{k} \gets \vx^{y}_{k} + \vU_{k} (\hvx^{(i)}_{k} - \vx^{y}_{k})$ \Comment{Pseudo Update}
            \vspace{2pt}
            \State $\vP^{f}_{k} \gets \vP^{y}_{k} - \vU_{k} \mathbf{\Delta}_{k} \vU_{k}^\top$
        \EndFor
        \vspace{2pt}
        \State Set $\hvx_{N} \gets \vx^{f}_{N}$ and $ \vP_{N} \gets \vP^{f}_{N}$
        \For{$k \gets N-1$ \textbf{to} $0$}
            \State $\vG_{k} \gets \vP^{f}_{k} \, \vF^\top_{k} \, (\vP^{p}_{k+1})^{-1}$
            \vspace{2pt}
            \State $\hvx_{k} \gets \vx^{f}_{k} + \vG_{k} (\hvx_{k+1} - \vx^{p}_{k+1})$ \Comment{Smoothing}
            \vspace{2pt}
            \State $\vP_{k} \gets \vP^{f}_{k} + \vG_{k} (\vP_{k+1} - \vP^{p}_{k+1}) \vG_{k}^\top$
        \EndFor
        \EndProcedure
    \end{algorithmic}
\end{algorithm}

\section{Implementation Strategies of Newton \\ Iterated Kalman Smoothers} \label{sec:ls_tr_newton}
In the upcoming sections, we describe two algorithms for a robust implementation of the Newton iterated Kalman smoother. The line-search and trust-region strategies that we incorporate into the Newton-IKS are realizations of fundamental principles in optimization for scaling and regularizing the update of an iterate $\hvx^{(i)}_{0:N}$ along a direction $\vp^{(i)}$ to guarantee a consistent reduction of the objective \cite{wright1999numerical}.

\subsection{Recursive Newton Method with Line Search} \label{sec:ls_Newton}
The procedure of line search assumes the existence of a direction $\vp^{(i)}$ at a current iterate $\vx^{(i)}_{0:N}$ and proposes an updated iterate $\vx^{(i+1)}_{0:N}$. The distance taken along the direction $\vp^{(i)}$ is scaled by a step size $\alpha > 0$ in a way that guarantees a reduction of the objective function
\begin{equation} \label{eq:line-search}
    \hvx^{(i+1)}_{0:N} = \hvx^{(i)}_{0:N} + \alpha \, \vp^{(i)}.
\end{equation}
In our case, the Newton-IKS from Section~\ref{sec:recursive-Newton-IKS} indirectly supplies the search direction of the smoothed trajectory via $\vp^{(i)} = \hvx_{0:N} - \hvx^{(i)}_{0:N}$, where $\hvx_{0:N}$ is the output of Algorithm~\ref{alg:Newton-IKS-additive} given the current iterate  $\hvx^{(i)}_{0:N}$ as a nominal trajectory.

However, the direction that the Newton-IKS delivers may not be a valid search direction as the Hessian of the objective function may not be positive-definite. 
To overcome this challenge, we propose a simple approach that increases the diagonal regularization factor $\lambda$ until the \emph{expected} cost reduction is positive $\tL(\hvx^{(i)}_{0:N}) - \tL(\hvx_{0:N}) > 0$ where $\tL(.)$ is the (regularized) second-order approximation in Equation~\eqref{eq:approx-objective-function}, which corresponds to a descent direction. 

Given a descent direction $\vp^{(i)}$, various approaches are available for choosing $\alpha$ exactly or approximately \cite{wright1999numerical}. We choose to apply a backtracking line-search scheme to find a step size $\alpha$ such that $L(\hvx^{(i)}_{0:N} + \alpha \, \vp^{(i)}) < L(\hvx^{(i)}_{0:N})$, where $L(.)$ is the \emph{original} nonlinear objective in Equation~\eqref{eq:objective-function}. Algorithm~\ref{alg:ls-Newton-IKS-additive} provides an overview of a Newton-IKS algorithm with an approximate line-search strategy.

\subsection{Recursive Newton Method with a Trust Region} \label{sec:trust-region}
While line-search techniques optimize the step size along a pre-defined search direction, trust-region methods intervene and directly modify the search direction based on an approximate model of the nonlinear objective in a region around the current iterate. The size of this region implies the relative trust of the local approximation and simultaneously influences both the update direction and the step size.

In the case of the Newton-IKS, we implement a trust-region technique akin to a Levenberg-Marquardt algorithm  \cite{madsen2008introduction}. This approach directly controls the regularization in Equation~\eqref{eq:approx-objective-function} to modify the search direction based on the quality of the local approximation. The quality is measured by the ratio of the \emph{actual} cost difference to the \emph{expected} cost difference given a nominal trajectory $\hvx^{(i)}_{0:N}$ and a candidate solution $\hvx_{0:N}$
\begin{equation*}
    \rho = \frac{\Delta L}{\Delta \tL} = \frac{L(\hvx^{(i)}_{0:N}) - L(\hvx_{0:N})}{\tL(\hvx^{(i)}_{0:N}) - \tL(\hvx_{0:N})}.
\end{equation*}
An update is accepted when $\rho > 0$, implying that the current approximation is close to the true underlying objective around the current iterate, and the trust region is enlarged accordingly by reducing $\lambda$. When $\rho \le 0$, the update is rejected, and the region is tightened by increasing $\lambda$. Algorithm~\ref{alg:tr-Newton-IKS-additive} provides an overview of the Newton-IKS with a trust-region strategy.

\begin{algorithm}
    \caption{Newton-IKS with Line Search}\label{alg:ls-Newton-IKS-additive}
    \begin{algorithmic}[1]
        \item \textbf{input:} Initial trajectory $\hvx^{(0)}_{0:N}$, measurements $\vy_{1:N}$,
        Models, Hessians, and Jacobians: $\vf, \vh, \vF_{\vx}, \vH_{\vx}, \vF_{\vx \vx}, \vH_{\vx \vx}$, constants: $\vm_0, \vP_0, \vQ, \vR$, backtracking mult. $\beta \in (0, 1)$, backtracking iterations $M$, and overall iterations $N_i$
        \vspace{2pt}
        \item \textbf{output:} The MAP trajectory $\hvx^{*}_{0:N}$
        \vspace{2pt}        
        \For{$0 \leq i < N_i$}
        \vspace{2pt}
        \State $\hvx_{0:N} \gets$ \textsc{Newton-IKS}($\hvx^{(i)}_{0:N}, \lambda=0$)
        \vspace{2pt}
        \If{$\tL(\hvx^{(i)}_{0:N}) - \tL(\hvx_{0:N}) > 0$}
            \vspace{2pt}
            \State $\vp^{(i)} \gets \hvx_{0:N} - \hvx^{(i)}_{0:N}$ \Comment{Descent direction}
        \Else
            \State Set $\lambda \gets 10^{-6}$
            \State $\hvx_{0:N} \gets$ \textsc{Newton-IKS}($\hvx^{(i)}_{0:N}, \lambda$) \Comment{Regularize}
            \vspace{2pt}
            \While{$\tL(\hvx^{(i)}_{0:N}) - \tL(\hvx_{0:N}) \leq 0$ \textbf{and} $\lambda \leq 10^{16}$}
                \State $ \lambda \gets 10 \, \lambda$
                \State $\hvx_{0:N} \gets$ \textsc{Newton-IKS}($\hvx^{(i)}_{0:N}, \lambda$)
                \vspace{2pt}
            \EndWhile
            \State $\vp^{(i)} \gets \hvx_{0:N} - \hvx^{(i)}_{0:N}$
        \EndIf
        \vspace{2pt}
        \State Set $\alpha \gets 1$, $m \gets 0$
        \vspace{2pt}
        \While{$L(\hvx^{(i)}_{0:N} + \alpha \, \vp^{(i)}) \geq L(\hvx^{(i)}_{0:N})$ \textbf{and} $m \leq M$}
            \vspace{2pt}
            \State $\alpha \gets \beta \, \alpha$, $m \gets m + 1$ \Comment {Backtracking}
        \EndWhile
        \vspace{2pt}
        \If{$L(\hvx^{(i)}_{0:N} + \alpha \, \vp^{(i)}) < L(\hvx^{(i)}_{0:N})$}
            \vspace{2pt}
            \State $\hvx^{(i+1)}_{0:N} \gets \hvx^{(i)}_{0:N} + \alpha \, \vp^{(i)}$ \Comment{Accept step}
        \Else
            \State $\hvx^{(i+1)}_{0:N} \gets \hvx^{(i)}_{0:N}$ \Comment{Reject step}
        \EndIf
        \vspace{2pt}
        \EndFor
    \end{algorithmic}
\end{algorithm}

\begin{algorithm}
    \caption{Newton-IKS with a Trust Region}\label{alg:tr-Newton-IKS-additive}
    \begin{algorithmic}[1]
        \item \textbf{input:} Initial trajectory $\hvx^{(0)}_{0:N}$, measurements $\vy_{1:N}$,
        Models, Hessians, and Jacobians: $\vf, \vh, \vF_{\vx}, \vH_{\vx}, \vF_{\vx \vx}, \vH_{\vx \vx}$, constants: $\vm_0, \, \vP_0, \, \vQ, \, \vR$, initial regularization $\lambda_0$, regularization mult. $\nu > 1$, and overall iterations $N_i$
        \vspace{2pt}
        \item \textbf{output:} The MAP trajectory $\hvx^{*}_{0:N}$
        \vspace{2pt}        
        \State Set $\lambda \gets \lambda_0$ $\nu \gets 2$
        \For{$0 \leq i < N_i$}
        \vspace{2pt}
        \State $\hvx_{0:N} \gets$ \textsc{Newton-IKS}($\hvx^{(i)}_{0:N}, \lambda$)
        \vspace{2pt}
        \State{$\Delta L \gets L(\hvx^{(i)}_{0:N}) - L(\hvx_{0:N})$} \Comment{Actual cost diff.}
        \vspace{2pt}
        \vspace{2pt}
        \State $\Delta \tL \gets \tL(\hvx^{(i)}_{0:N}) - \tL(\hvx_{0:N})$ \Comment{Expected cost diff.}
        \vspace{2pt}
        \vspace{2pt}
        \State{$\rho \gets \Delta L / \Delta \tL$}
        \vspace{2pt}
        \If{$\rho > 0$ \textbf{and} $\Delta \tL > 0$}
        \vspace{2pt}
        \State $\lambda \gets \lambda \, \text{max}\{\frac{1}{3}, 1 - (2 \, \rho - 1)^3\}$, $\nu \gets 2$
        \State $\hvx^{(i+1)}_{0:N} \gets \hvx_{0:N}$ \Comment{Accept step}
        \Else
        \State $\lambda \gets \nu \, \lambda$, $\nu \gets 2 \, \nu$
        \vspace{2pt}
        \State $\hvx^{(i+1)}_{0:N} \gets \hvx^{(i)}_{0:N}$ \Comment{Reject step}
        \vspace{2pt}
        \EndIf
        \EndFor
    \end{algorithmic}
\end{algorithm}

\clearpage

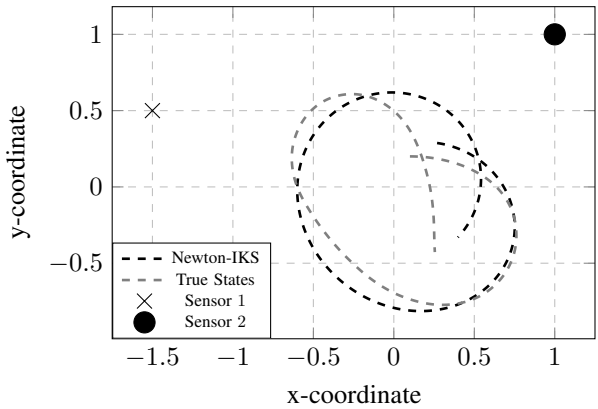
\begin{figure}[H]
    \centering
    \input{figures/xyplot.tex}
    \vspace{-0.25cm}
    \caption{Example of a smoothed trajectory obtained from a Newton-IKS with a trust-region method in the coordinated turn model.}
    \label{fig:xyplot}
\end{figure}

\section{Experimental Results} \label{sec:experiment}
In this section, we assess the performance of the proposed approaches using a simulated coordinated turn model example with bearings-only measurements \cite{sarkka2020levenberg, skoglund2015extended, fatemi2012study}. The system has a 5-dimensional state vector $\vx = [p_x, \, p_y, \, \dot{p}_x, \, \dot{p}_y, \, \omega]^\top$ which describes the $x-y$ position, the $x-y$ velocity, and the turn rate of the target. The bearing is measured by two sensors located at known positions. Figure \ref{fig:xyplot} depicts an example true trajectory, an estimated trajectory using a trust-region Newton-IKS, and the locations of the two sensors. 

In addition to our recursive algorithms, we implement the equivalent batch optimization techniques as presented in \cite{wright1999numerical} and use the same hyperparameters in the line-search and trust-region variants. We focus on comparing the computational complexity of the recursive and batch techniques. We rely on JAX~\cite{jax2018github} for automatic differentiation.

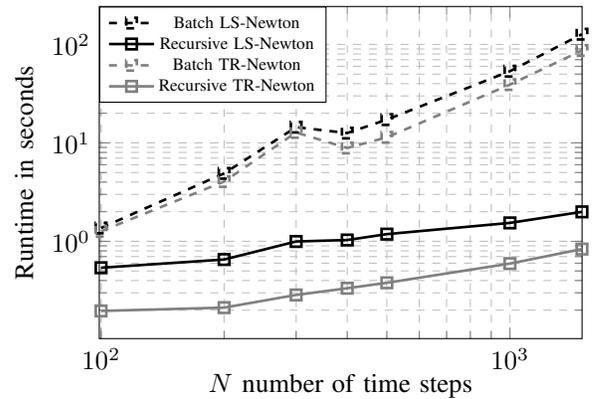
\begin{figure}[!t]
    \centering
    \input{figures/runtime.tex}
    \vspace{-0.25cm}
    \caption{Runtime comparison of the batch Newton method against the recursive trust-region (TR) and line-search (LS) Newton algorithms.}
    \label{fig:runtime}
    \vspace{-0.35cm}
\end{figure}

In this study, we investigate trajectories of different lengths, ranging from $N=100$ to $N=1500$, and report the average runtime (over $20$ runs) of running $30$ overall iterations of the iterated batch and recursive Newton methods. The average runtime as a function of trajectory length is illustrated in Figure ~\ref{fig:runtime}. As expected, the computational performance of the recursive Newton algorithms is superior to their batch counterpart in terms of runtime. An open-source implementation is available at \url{https://github.com/hanyas/second-order-smoothers}.

\section{Conclusion}
We presented a computationally efficient realization of Newton's method for smoothing in nonlinear state-space models with additive noise. We leveraged automatic differentiation tools to compute the required first- and second-order derivatives with minimal effort and formulated a corresponding affine state-space model with augmented pseudo measurements. We showed that this modified SSM form enables the implementation of a recursive, computationally favorable Kalman smoothing algorithm equivalent to a Newton step. Furthermore, We proposed line-search and trust-region extensions of the proposed method to ensure the convergence to a local optimum. Finally, we empirically validated the efficiency of our recursive Newton method against standard batch solutions.

\bibliographystyle{IEEEtran}
\bibliography{references}

\end{document}

%% file: figures/xyplot.tex
\begin{tikzpicture}
\begin{axis}[
    width=8cm,
    height=6cm,
    legend style={
        nodes={scale=0.65, transform shape},
        at={(0,0)},
        anchor=south west
    },
    grid=both,
    grid style=dashed,
    xlabel={x-coordinate},
    ylabel={y-coordinate},
    ]
\addplot[black, dashed, line width=1pt] table [x=px, y=py, col sep=comma]{figures/xyplot.csv};
\addplot[gray, dashed, line width=1pt] table [x=xtrue, y=ytrue, col sep=comma]{figures/xyplot.csv};
\addplot [only marks, mark=x, mark size=4pt,  draw=black] coordinates { (-1.5, 0.5) };
\addplot [only marks, mark=*, mark size=4pt, fill=black,  draw=black] coordinates { (1, 1) };
\legend{Newton-IKS, True States, Sensor $1$, Sensor $2$}
\end{axis}
\end{tikzpicture}

%% file: figures/runtime.tex
\begin{tikzpicture}
\begin{axis}[
    width=8cm,
    height=6cm,
    legend style={
        nodes={scale=0.65, transform shape},
        at={(0,1)},
        anchor=north west
    },
    grid=both,
    grid style=dashed,
    xmin=99, xmax=1501,
    xlabel={$N$ number of time steps},
    ylabel={Runtime in seconds},
    ymode=log,
    xmode=log,
    xlabel style={yshift=5pt}
    ]
\addplot[black, dashed,mark=square, line width=1pt] table [x=times, y=cpu_ls_batch_runtime, col sep=comma]{figures/ls_batch_seq_runtime.csv};
\addplot[black,mark=square, line width=1pt] table [x=times, y=cpu_ls_recursive_runtime, col sep=comma]{figures/ls_batch_seq_runtime.csv};
\addplot[gray, dashed,mark=square, line width=1pt] table [x=times, y=cpu_tr_batch_runtime, col sep=comma]{figures/tr_batch_seq_runtime.csv};
\addplot[gray,mark=square, line width=1pt] table [x=times, y=cpu_tr_recursive_runtime, col sep=comma]{figures/tr_batch_seq_runtime.csv};
\legend{Batch LS-Newton, Recursive LS-Newton, Batch TR-Newton, Recursive TR-Newton}
\end{axis}
\end{tikzpicture}

%% file: main.bbl
\begin{thebibliography}{10}
\providecommand{\url}[1]{#1}
\csname url@samestyle\endcsname
\providecommand{\newblock}{\relax}
\providecommand{\bibinfo}[2]{#2}
\providecommand{\BIBentrySTDinterwordspacing}{\spaceskip=0pt\relax}
\providecommand{\BIBentryALTinterwordstretchfactor}{4}
\providecommand{\BIBentryALTinterwordspacing}{\spaceskip=\fontdimen2\font plus
\BIBentryALTinterwordstretchfactor\fontdimen3\font minus
  \fontdimen4\font\relax}
\providecommand{\BIBforeignlanguage}[2]{{%
\expandafter\ifx\csname l@#1\endcsname\relax
\typeout{** WARNING: IEEEtran.bst: No hyphenation pattern has been}%
\typeout{** loaded for the language `#1'. Using the pattern for}%
\typeout{** the default language instead.}%
\else
\language=\csname l@#1\endcsname
\fi
#2}}
\providecommand{\BIBdecl}{\relax}
\BIBdecl

\bibitem{sarkka2013bayesian}
S.~S{\"a}rkk{\"a}, \emph{Bayesian Filtering and Smoothing}.\hskip 1em plus
  0.5em minus 0.4em\relax Cambridge University Press, 2013.

\bibitem{Bar-Shalom+Li+Kirubarajan:2001}
Y.~Bar-Shalom, X.-R. Li, and T.~Kirubarajan, \emph{Estimation with Applications
  to Tracking and Navigation}.\hskip 1em plus 0.5em minus 0.4em\relax Wiley,
  2001.

\bibitem{bar1995multitarget}
Y.~Bar-Shalom and X.-R. Li, \emph{Multitarget-Multisensor Tracking:
  {P}rinciples and Techniques}.\hskip 1em plus 0.5em minus 0.4em\relax Yaakov
  Bar-Shalom, 1995.

\bibitem{kalman1960new}
R.~E. Kalman, ``A new approach to linear filtering and prediction problems,''
  \emph{Transactions of the {ASME} journal of Basic Engineering}, 1960.

\bibitem{rauch1965maximum}
H.~E. Rauch, F.~Tung, and C.~T. Striebel, ``Maximum likelihood estimates of
  linear dynamic systems,'' \emph{AIAA Journal}, 1965.

\bibitem{jazwinski2007stochastic}
A.~H. Jazwinski, \emph{Stochastic Processes and Filtering Theory}.\hskip 1em
  plus 0.5em minus 0.4em\relax Academic Press, 1970.

\bibitem{julier2000new}
S.~Julier, J.~Uhlmann, and H.~F. Durrant-Whyte, ``A new method for the
  nonlinear transformation of means and covariances in filters and
  estimators,'' \emph{Transactions on Automatic Control}, 2000.

\bibitem{bell1994iterated}
B.~M. Bell, ``The iterated {K}alman smoother as a {G}auss--{N}ewton method,''
  \emph{SIAM Journal on Optimization}, 1994.

\bibitem{Garcia+Svensson+Moreland+Sarkka:2015}
{\'A}.~F. Garc\'ia-Fern\'andez, L.~Svensson, M.~Morelande, and
  S.~S{\"a}rkk{\"a}, ``Posterior linearization filter: {P}rinciples and
  implementation using sigma points,'' \emph{IEEE Transactions on Signal
  Processing}, 2015.

\bibitem{Garcia:2017}
{\'A}.~F. Garc{\'i}a-Fern\'andez, L.~Svensson, and S.~S{\"a}rkk{\"a},
  ``Iterated posterior linearization smoother,'' \emph{IEEE Transactions on
  Automatic Control}, 2017.

\bibitem{Tronarp+Fernandez+Sarkka:2018}
F.~Tronarp, A.~F. Garc\'ia-Fern\'andez, and S.~S\"arkk\"a, ``Iterative
  filtering and smoothing in nonlinear and non-{G}aussian systems using
  conditional moments,'' \emph{IEEE Signal Processing Letters}, 2018.

\bibitem{sarkka2020levenberg}
S.~S{\"a}rkk{\"a} and L.~Svensson, ``{L}evenberg-{M}arquardt and line-search
  extended {K}alman smoothers,'' in \emph{International Conference on
  Acoustics, Speech and Signal Processing}.\hskip 1em plus 0.5em minus
  0.4em\relax IEEE, 2020.

\bibitem{wright1999numerical}
S.~Wright and J.~Nocedal, \emph{Numerical Optimization}.\hskip 1em plus 0.5em
  minus 0.4em\relax Springer, 1999.

\bibitem{garcia2014gaussian}
{\'A}.~F. Garc{\'\i}a-Fern{\'a}ndez and L.~Svensson, ``Gaussian {MAP} filtering
  using {K}alman optimization,'' \emph{Transactions on Automatic Control},
  2014.

\bibitem{fatemi2012study}
M.~Fatemi, L.~Svensson, L.~Hammarstrand, and M.~Morelande, ``A study of {MAP}
  estimation techniques for nonlinear filtering,'' in \emph{International
  Conference on Information Fusion}.\hskip 1em plus 0.5em minus 0.4em\relax
  IEEE, 2012.

\bibitem{moriyama2003incremental}
H.~Moriyama, N.~Yamashita, and M.~Fukushima, ``The incremental {G}auss-{N}ewton
  algorithm with adaptive stepsize rule,'' \emph{Computational Optimization and
  Applications}, 2003.

\bibitem{Bell:1993}
B.~M. Bell and F.~W. Cathey, ``The iterated {K}alman filter update as a
  {G}auss--{N}ewton method,'' \emph{Transactions on Automatic Control}, 1993.

\bibitem{humpherys2012fresh}
J.~Humpherys, P.~Redd, and J.~West, ``A fresh look at the {K}alman filter,''
  \emph{SIAM Review}, 2012.

\bibitem{skoglund2015extended}
M.~A. Skoglund, G.~Hendeby, and D.~Axehill, ``Extended {K}alman filter
  modifications based on an optimization view point,'' in \emph{International
  Conference on Information Fusion}.\hskip 1em plus 0.5em minus 0.4em\relax
  IEEE, 2015.

\bibitem{ollivier2019extended}
Y.~Ollivier, ``The extended {K}alman filter is a natural gradient descent in
  trajectory space,'' \emph{arXiv preprint arXiv:1901.00696}, 2019.

\bibitem{madsen2008introduction}
K.~Madsen and H.~B. Nielsen, \emph{Introduction to Optimization and Data
  Fitting}, 2008.

\bibitem{jax2018github}
J.~Bradbury, R.~Frostig, P.~Hawkins, M.~J. Johnson, C.~Leary, D.~Maclaurin, and
  S.~Wanderman-Milne, ``{{JAX}: {C}omposable transformations of
  {P}ython+{N}um{P}y programs},'' \url{http://github.com/google/jax}, 2018.

\end{thebibliography}
